\documentclass[%
reprint,
 showkeys,
 amsmath,amssymb,
 aps,
]{revtex4-1}

\usepackage{graphicx}
\usepackage{dcolumn}
\usepackage{bm}
\usepackage{lineno}

\begin{document}


\title{Gamilaraay kinship revisited: incidence of recessive disease is dynamically traded-off against benefits of cooperative behaviours}
\author{Jared M. Field}
  \email{jared.field@unimelb.edu.au}
\affiliation{%
School of Mathematics and Statistics, University of Melbourne, Melbourne, Australia\\
 }%


\date{\today}

\begin{abstract}
Traditional Indigenous marriage rules have been studied extensively since the mid 1800s. Despite this, they have historically been cast aside as having very little utility. This is, in large part, due to a focus on trying to  understand broad-stroke marriage restrictions or how they may evolve. Here, taking the Gamilaraay system as a case study, we instead ask how relatedness may be distributed under such a system. We show, remarkably, that this system dynamically trades off kin avoidance to minimise incidence of recessive diseases against expected levels of cooperation, as understood formally through Hamilton’s rule.
\end{abstract}




\maketitle

\section*{Introduction}
In his hugely influential work Ancient Society Lewis Morgan wrote that the Gamilaraay kinship system  represents `the most primitive form of society hitherto discovered'\cite{morgan1877ancient}. Such sentiment persisted for decades, with many academics reluctant to throw out old ideas of Social Evolutionism, whereby all societies were perceived to be on a sliding scale of savagery to civilisation (see Chapter 2 of \cite{mcconvell2018skin} for a survey of such ideas in the Australian context). More rigorous (and slightly less colonial) work arrived with the seminal study of Levi-Strauss \cite{levi2020structures}. Here, for example, the mathematician André Weil used group theory to understand precisely who can marry whom in kinship systems, such as that of the Gamilaraay, where descent operates in cycles. Since then most work has followed Weil to attempt to understand broad-stroke marriage restrictions or indeed how they might come about, with studies in this area more generally remaining hotly debated (see for example \cite{itao2020evolution, read2020simulation, itao2020reply}). Here, we take a different tact entirely and instead ask how relatedness may be distributed in such societies. More than this, we ask how might this distribution change over time. In this way, we aim to better grasp not just the form of the Gamilaraay kinship system but also its function.

To this end, we worked out the exact distribution of common ancestors that may appear in a general Gamilaraay pedigree. Coupled with kinship coefficients, this allowed for the derivation of an average kinship coefficient for a couple permitted to reproduce in this system. Under the assumption that inbreeding coefficients are negligible at the great grandparent stage, we found that the entire Gamilaraay nation would need to reduce to just 24 individuals so that couples may (on average) be as closely related as full first cousins. While striking, inbreeding coefficients may of course not always be negligible. For this reason we dropped this assumption, and instead asked how average kinship coefficients change over time. We found that there is, in fact, an equilibrium point towards which average kinship is pushed. This point is a function of reproductive population size, and so dynamically re-calibrates as that quantity changes. In this way, average kinship may both decrease and increase in reaction to historical values. Remarkably, the Gamilaraay kinship system dynamically trades-off the incidence of recessive diseases against the benefits of cooperative behaviours, formally understood through Hamilton's Rule \cite{frank2013naturalhist}. Otherwise put, the Gamilaraay kinship system does not simply strive to create kin restrictions, but instead has an underlying utility function that also accounts for social benefits. Contrary to the beliefs of Morgan, Gamilaraay culture appears to be anything but primitive.  

The rest of this paper is organised as follows: we start by providing background on the Gamilaraay system. Following this, we map a general pedigree by sub-types, or marriageable groups, to show where uncertainty may first enter a family tree. Next, we use combinatoric arguments to derive explicit distributions of common ancestors first on a particular branch, and then across all branches at the great grandparent stage. In turn, we use this to derive average kinship coefficients for a general permitted couple and consider their dynamics. Finally, we contextualise our findings in the current theoretical biology literature, discuss their consequences and consider future directions of work.

\section*{Gamilaraay System}
In the Gamilaraay kinship system, as with many others of this region (see for example \cite{mcconvell2018skin} for a more linguistic treatment),  all individuals are divided into two groups, namely Guwaymadhaan (Dark or Heavy Blood) and Guwaygaliyarr (Light Blood). Each child inherits belonging to a particular group directly from their mother, so that, for example, a Guwaymadhaan mother produces a Guwaymadhaan child. These two groups are split further in two so that, for the purpose of marriage, there are four subgroups in total. These subgroupings are such that a child belongs to the subgroup opposite to that of their mother. In the Gamilaraay system, for example, a Yibadhaa woman produces Gambuu sons and Buudhaa daughters (while the language is different, due to gender, the types are the same.). If a Buudhaa woman reproduces, her children then once again become Yibaay/Yibadhaa, which will not however be the case for her Gambuu brother. Marriage rules are then such that one can only marry an individual that is of opposite blood type, and opposite subgrouping to their father. This is presented more clearly in Fig. 1. Here general labels are adopted both to ease mathematical analysis later and because this system extends beyond the Gamilaraay nation. See \cite{fison1880kamilaroi,parker1905euahlayi} for older anthropological treatments of the Gamilaraay system and \cite{Spearim2014a, Spearim2014b} for a modern primary source account.  A few obvious kin marriage restrictions, from Fig. 1, become clear. For example, an individual may not marry siblings, any maternal cousins related through females, nor any aunties or uncles. Paternal cousins are possible (likely half, due to marriage norms), but as we will show, the probability is low.

\begin{figure}

\caption{Kinship dynamics of the four group Gamilaraay system. Solid lines represent matrilineal inheritance of subgrouping whereas dashed lines link subgroups that are permitted to marry. Individuals are only permitted to marry someone of the opposite bloody type (CD type people, if an individual is of type AB, for example) and opposite subgrouping to their father (C type people, if an individual is B type). Note that this ensures that an individual can only marry someone who is of type different to themselves, and both their parents. By virtue of inheritance rules it also ensures individuals do not marry siblings, aunties, uncles and maternal cousins of any order related through females. 
}

\includegraphics[scale=4]{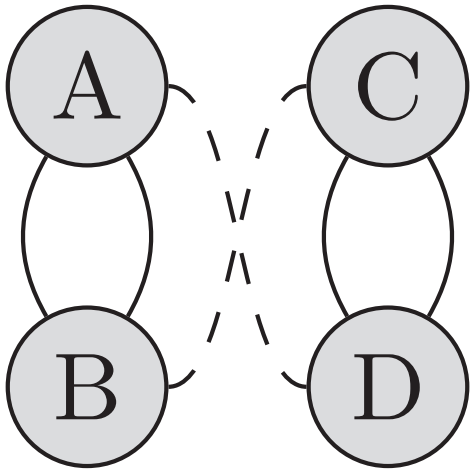}
\label{fig:Fig1}
\end{figure} 

\section*{General Pedigree}
 To work out the average relatedness of a couple, first note that by virtue of this system, uncertainty enters pedigrees only at the great-grandparent stage. This is most clear in Fig. 2. Here, we take a general individual of subgrouping type B, and trace their pedigree. This individual will have a mother of type A and father of type D (denoting movement to the left and right as female and male ancestors, respectively). Tracing back one generation further, a type A individual will have a type B mother and C father whereas a type D individual will have a type C mother and B father. While there are two Bs and two Cs at this stage, they must be distinct individuals owing to their different sexes. Repeating this process, we see that at the great-grandparent stage there will be four types As and four type Bs. However, they will be split across the sexes. In other words, there will be precisely two male As, two females As, two male Ds, and two female Ds. At this stage, then, it is now possible that one individual may occupy two distinct spots in this genealogical tree. In the next section, we work out the precise probability of one of those spots being filled by the same or distinct individuals. 

\begin{figure}[h!]

\caption{General pedigree of an individual in the Gamilaraay system following the subgrouping notation as in Fig. 1. Sex in this figure is encoded by left (female) and right (male) branches.  While this pedigree is presented for an individual of type B, due to symmetry, this general form applies to all individuals. From Fig. 1. we see that a type B individual will have a type A mother and type D father. In turn, a type A individual will have a type B mother and type C father. A type D individual, however, will have a type C mother and type B father. Note that uncertainty only enters at the great-grandparent stage, where one individual may occupy two distinct spots in this potential geneological tree. For example, one type A female (left branches) may potentially fill both type A female positions (first and seventh). This will make her the same mother to the type B female and male in the generation below, and hence increase the relatedness of the AD couple one generation later.
}

\includegraphics[scale=2]{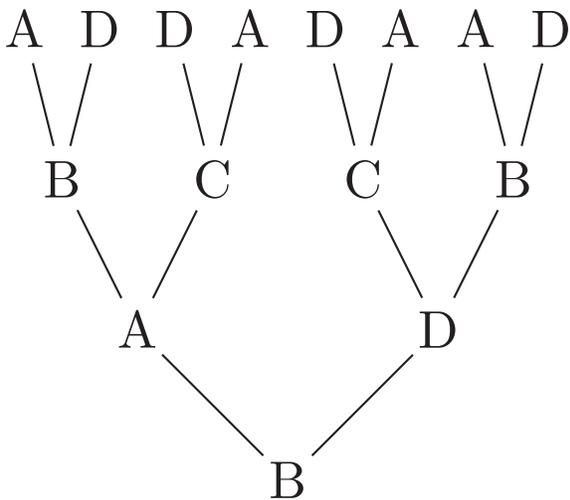}
\label{fig:Fig2}
\end{figure} 


\section*{Distribution of common ancestors on one branch}
Suppose there are $n$ reproductive individuals in the population belonging to our subgroup and sex of interest. Further, suppose we assign each of these individuals a unique label between $1,...,n$. With this notation, we may enumerate every possible combination of individuals (to fill the two permitted genealogical spots) by multisets of the form $\{1,1\}$, $\{1,2\}$, $\{1,3\}$, ... , $\{n-1, n\}$, $\{n,n\}$. $\{1,1\}$, for example, represents the instance where the first individual occupies both spots. $\{6,1\}$ represents the instance where the sixth and first individual occupy a spot each, and so forth. The total number of different ways this section of the pedigree may be realised will then be given by the total number of multisets of the above form. This, in turn, is well known to be given by the binomial coefficient: 
\begin{equation}
   { n+1 \choose 2}.
\end{equation}
The number of instances where an individual occupies both spots (which is to say, of the form $\{3,3\}$, for example) is simply the total number of individuals $n$. Assuming each instance is equally likely, this in turn means that the probability that both spots are occupied by the same individual in general will be:
\begin{align}
    p_{common}&= \frac{n}{ { n+1 \choose 2}},\\
    &= \frac{2}{ n+1 }. \label{common}
\end{align}
If the two spots are occupied by different individuals, the first may be filled by $n$ individuals, while the second may be filled by the remaining $(n-1)$. As we do not care about order, we need to factor out repeats by $2!$ (the ordered couple $(1,2)$ for our purposes is equivalent to $(2,1)$). Hence, the number of instances where individuals are distinct will be given by:
\begin{equation}
    \frac{n\cdot (n-1)}{2!},
\end{equation}
so that the probability the two spots are filled by distinct individuals is
\begin{align}
  p_{distinct} &= \frac{ \frac{n\cdot (n-1)}{2!}}{ { n+1 \choose 2}}, \\
  &=  \frac{n-1}{n+1}.
\end{align}

\section*{Distribution of common ancestors across all branches}
Recall that at the great-grandparent stage there are eight individuals in total, spread across the two sexes and four subgroups. In the previous section, we worked out the probability that two permitted spots of a particular subtype are filled by the same or different individuals. We now need to combine these probabilities to work out the distribution of any number of common ancestors at this generation. In other words, the probability that there are $4, 6,7$ or $8$ distinct individuals in this generation in total.

To do so, note that there are four subtypes in this generation: male and female type As, and male and female type Ds, for this pedigree. If we encode the event that one individual occupies both genealogical spots of a subtype by $1$ and the other case by $2$ then we can use 4-tuples to count ancestors. For example, the 4-tuple $(1,2,2,2)$ represents the instance where the first subtype has one individual in both spots, while the other subtypes have distinct individuals. In other words, of the total possible eight ancestors at this stage there are only seven. In turn this would mean that our couple of interest in Fig. 2, individuals A and D in generation one, will be related. With this notation, it is clear that the total number of combinations where there are seven ancestors will be given by the binomial coefficient $4\choose 1$, so that the probability of having seven ancestors at the great grandparent stage is 
\begin{equation}
    p_7 = {4\choose 1}\cdot (p_{common})\cdot (p_{distinct})^3.
\end{equation}
With similar reasoning the other probabilities can be shown to be given by
\begin{align}
    p_4 & = {4\choose 4}\cdot (p_{common})^4, \\
%
    p_5 & = {4\choose 3}\cdot (p_{common})^3\cdot (p_{distinct}),\\
    p_6 & = {4\choose 2}\cdot (p_{common})^2\cdot (p_{distinct})^2,\\
    p_8 & = {4\choose 0}\cdot  (p_{distinct})^4.
\end{align}

\section*{Average kinship coefficient}
If there were no uncertainty in our problem, and we knew exactly how many common ancestors were at a given stage, calculating coefficients of kinship, a measure of relatedness useful in the study of disease, would be simple. For example, suppose there were one common ancestor at the great grandparent stage, a type D female. This would mean that one copy of the tree as per Fig. 3. would appear in this particular pedigree. To work out the coefficient of kinship of the AD couple, we would simply need to count the number of individuals in the path between them, and raise $1/2$ to that number. In this case, assuming that their common ancestor D is not the result of related individuals, their coefficient would be $1/2^5$. If instead their common ancestor D is in turn the offspring of related individuals, their coefficient of kinship would be accordingly higher. In particular, we can write their coefficient of kinship as $1/2^5 \cdot (1+f_D)$, where $f_D$ is the coefficient of inbreeding of D \cite{wright1922coefficients, crow1986basic}.

More generally, if there are $k$ common ancestors at this stage then the coefficient of kinship of the $AD$ couple can be written as 

\begin{equation}
   f_{AD} = \sum^k_{i=1} \frac{1}{2^5}(1+f_i),
\end{equation}
where $f_i$ is the coefficient of inbreeding of each of the $k$ common ancestors. Averaging over the distribution that the $f_i$ are drawn from we can write 
\begin{align}
    \bar f_{AD} &=\sum^k_{i=1} \frac{1}{2^5}(1+\bar f),\\
    &= \frac{1}{2^5}(k+k\bar f).
\end{align}
There is also however uncertainty in the total number of common ancestors $k$. Averaging over $k$, using the probabilities from the previous section, we obtain the average coefficient of kinship:

\begin{align}
    \bar{\bar{f}}_{AD} &= \sum^4_{k=0}
    \frac{1}{2^5}(k+k\bar f)p_{8-k},\\
    &= \frac{1}{2^5}(\bar k + \bar k \bar f), \label{AD}
\end{align}
where bars indicate averages. 


In Fig. 4, we plot $\bar{\bar{f}}_{AD}$ as a function of $n$, one eighth of the total number of reproductive individuals in the population. Here we assume $\bar{f} = 0$, an assumption we drop later. References lines are added to show relatedness of full first cousins (top grey) and full second cousins (bottom grey). Observe that the total population would need to reduce to $24$ individuals $(n=3)$ for average relatedness of couples to be the same as full first cousins.
\begin{figure}

\caption{Section of a pedigree assuming one common ancestor at the great grandparent stage. To calculate the kinship coefficient of the couple AD, we need to count the number of individuals in the paths between them (5) and raise $1/2$ to that number. Here, $f=1/2^5$.  }

\includegraphics[scale=3]{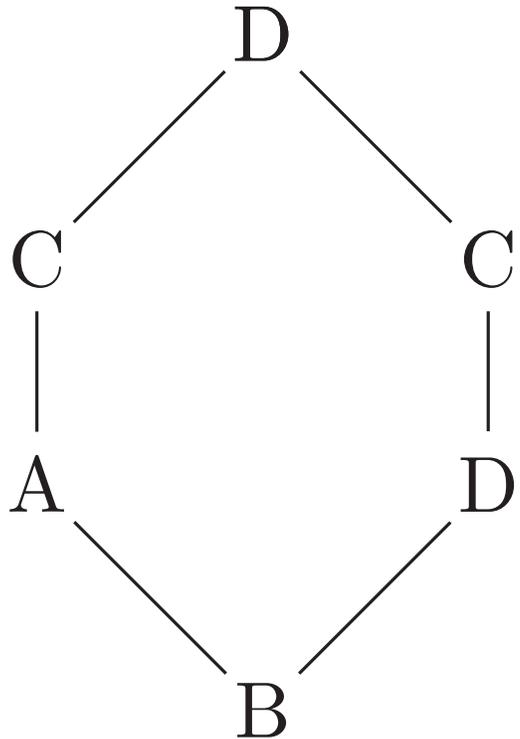}
\label{fig:Fig3}
\end{figure} 

\begin{figure}
\centering
\caption{Average coefficient of kinship $\bar{\bar{f}}$ of a randomly sampled couple operating under the Gamilaraay system as a function of $n$, one eighth of the total reproductive population size. Here we assume $\bar{f}$, inbreeding at the great grandparent stage, is zero (an assumption we drop later). Grey reference lines indicate  $\bar{\bar{f}}$ for full first (top) and full second (bottom) cousins. Note that, for a couple to be related as closely (on average) as full first cousins, the total population size needs to reduce to $24$ individuals ($n=3$).}

\includegraphics[scale=0.5]{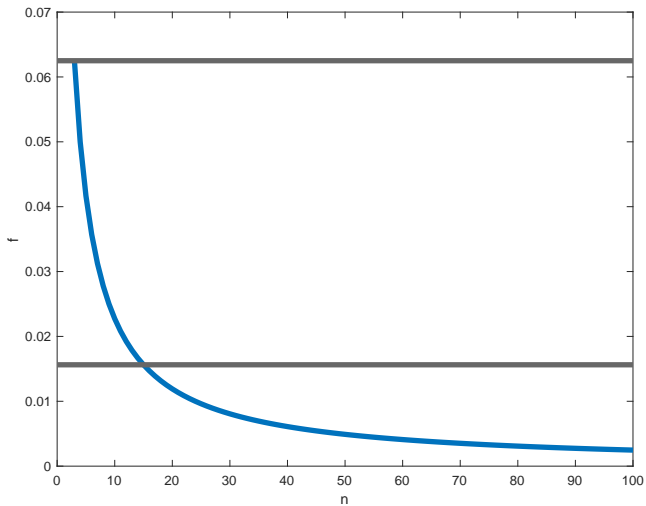}
\label{fig:Fig4}
\end{figure}
\section*{Dynamics}
While in the previous section we calculated the average kinship coefficient for an $AD$ couple, our equation holds for any permitted mating. To derive the equivalent equation for a $BC$ couple, we would simply need to reseed the original pedigree with either an $A$ or $D$ child. In light of this, we can drop the $AD$ from equation \eqref{AD} to write
\begin{equation}
     \bar{\bar{f}} 
    = \frac{1}{2^5}(\bar k + \bar k \bar f), \label{ave}
\end{equation}
which is the average kinship coefficient of any couple operating under the Gamilaraay system. To understand how kinship coefficients may change over time, note that the coefficient of inbreeding of an individual $f_i$, is in fact the coefficient of kinship of their parents (see \cite{crow1986basic}). In turn, this means the average coefficient of inbreeding is the average coefficient of kinship in the generation prior. With this definition, we can introduce a new index $j$ that tracks generations and replace the inbreeding coefficient with the kinship coefficient one generation earlier to rewrite \eqref{ave} as:
\begin{equation}
      \bar{\bar{f}}_j 
    = \frac{1}{2^5}(\bar k + \bar k \bar{\bar{f}}_{j-1} ),
\end{equation}
where $\bar{\bar{f}}_j $ and $\bar{\bar{f}}_{j-1} $ are the average kinship coefficients at the current and great great grandparent stage.

The change over generations can then be written as 
\begin{align}
   \bar{\bar{f_j}}-  \bar{\bar{f}}_{j-1} &= \frac{1}{2^5}(\bar k + (\bar k-2^5) \cdot \bar{\bar{f}}_{j-1})\\
   &= \frac{\bar k}{2^5}(1 + ( 1-\frac{2^5}{\bar k}) \cdot \bar{\bar{f}}_{j-1}), \label{positive}
\end{align}
where \eqref{positive} simply factors out positive terms (note that $\bar k $ approaches zero only as $n$ approaches infinity). In turn, this means the sign of $\bar{\bar{f_j}}-  \bar{\bar{f}}_{j-1}$ is fully determined by the other factor. In particular, the inequality
\begin{equation}
 \bar{\bar{f_j}}-  \bar{\bar{f}}_{j-1}<0
\end{equation}
will hold precisely when 
\begin{equation}
    1 + ( 1-\frac{2^5}{\bar k} ) \cdot \bar{\bar{f}}_{j-1}< 0.
\end{equation}
In turn, noting that $ 2^5/ \bar k$ is always greater than unity, this is equivalent to 
 \begin{equation}
        \bar{\bar{f}}_{j-1} \ge \frac{-1}{1-\frac{2^5}{\bar k}}:= f_{critical}.
    \end{equation}
In other words, if in the past the average kinship coefficient is greater than this $f_{critical}$ then the future average kinship will be smaller. The converse is also true. Note that $f_{critical}$, due to $\bar k$, is a function of $n$. In this way, if the population is neither growing nor decreasing then $f_{critical}$ is an equilibrium. If $n$ is changing (perhaps, for example, due to drought), then so too will $f_{critical}$. In this way, the point towards which average kinship is pushed is in fact dynamic. This is shown in Fig. 5 where the average kinship coefficient $\bar{\bar{f}}$ is plotted alongside $f_{critical}$, both as functions of $n$. Note that, for large $n$, $f_{critical}$ and $\bar{\bar{f}}$ coincide. This is not true when $n$ is small. This means that for small $n$ the population is pushed toward a state that resembles one in which ancestors were slightly related. Most importantly, however, observe that average kinship coefficients may both increase and decrease in reaction to historical values, and changes in $n$. 

\section*{Costs and benefits}
High kinship coefficients, and so too high inbreeding coefficients (a zygote may be viewed as the sampling of genes from parents), are well known to incur costs arising from increased incidence of recessive diseases \cite{crow1986basic}. In this way, the value of decreasing $\bar{\bar{f}}$ is immediately clear. The reason why it may be desirable for a system to sometimes tend to increase $\bar{\bar{f}}$ is less obvious. Note, however, that the coefficient of kinship we have calculated is two times the coefficient of relatedness \cite{crow1986basic}, a quantity that is critical to our understanding of the evolution of altruism \cite{gardner2011genetical}. More specifically, Hamilton's Rule tells us that we may expect a costly altruistic act to spread through a population if the benefit to others, weighted by this relatedness, outweighs the costs to the self \cite{frank2013naturalhist}. An increase in $\bar{\bar{f}}$, then, we should expect to also lead to an increase in cooperation, as understood through social evolution theory \cite{gardner2011genetical}. In turn this means that the Gamilaraay system dynamically trades off the costs of being too closely related against the benefits of cooperative behaviours.
    \begin{figure}
\centering
\caption{Average coefficient of kinship $\bar{\bar{f}}$ of a randomly sampled couple operating under the Gamilaraay system as a function of $n$, one eighth of the total reproductive population size. Overlayed in pink is $f_{critical}$, above (below) which future average kinship will decrease (increase). If the reproductive population size is stable then $f_{critical}$ is a true equilibrium, otherwise it recalibrates with changes in $n$.}

\includegraphics[scale=0.5]{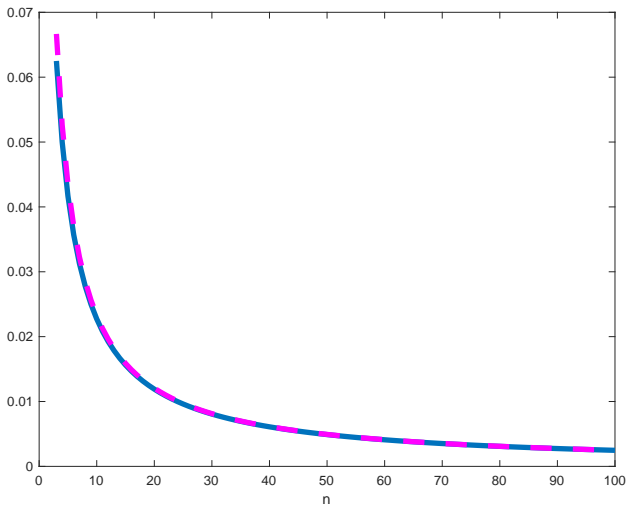}
\label{fig:Fig4}
\end{figure}
\section*{Discussion}
Colonial expansion of what was to become Australia, from around the mid-1800s, led to a flurry of anthropological studies of its original inhabitants \cite{taplin1878narrinyeri, fison1880kamilaroi,mathews1898victorian,mathews1900divisions,mathews1901ethnological, parker1905euahlayi, brown1918notes}. These works, if nothing else, provided a broad stroke understanding of kinship on the continent. A more rigorous study came only decades later with the seminal work of Levi-Strauss \cite{levi2020structures}. Indeed, in the appendix to this work, the mathematician André Weil of the Bourbaki set used group theory to provide deeper insight into precisely who can marry whom in these sorts of systems \cite{levi2020structures}. Mathematical interest, no doubt, was sparked by the oddity of cycles present in rules of descent typical to the continent. Work steadily increased throughout the following century (see for example \cite{hiatt1967authority, maddock1969alliance, hiatt1996arguments}), with studies on the resulting incest taboos remaining hotly debated \cite{itao2020evolution, read2020simulation, itao2020reply}. All of these studies, however, attempt to determine broad kin restrictions. In other words, following Weil, who can reproduce with whom? Here, instead, we ask how relatedness will be distributed in a society operating under such kinship systems. In particular, we take the Gamilaraay kinship system as a case study and ask: how related on average will a permitted couple be? Further, how might this average relatedness change over time?

To this end, we started by observing that the Gamilaraay kinship system categorises all individuals into four distinct subtypes. 
We then noted that, coupled with descent cycles, uncertainty in pedigrees enters the Gamilaraay system only at the great grandparent stage, as in Fig. 2. A consequence of this is that it is only then that two genealogical spots in a potential Gamilaraay family tree may be occupied by the same individual. Following this, we worked out the precise probability that such an event is realised, as a function of $n$, the number of reproductive individuals in a subtype of interest. With this, it was then possible to write down the exact distribution of common or distinct ancestors for a general pedigree at this stage. We then linked this distribution with kinship coefficients (employed often in the study of disease \cite{wright1922coefficients, crow1986basic}) to derive an average kinship coefficient for a general couple operating under this system. 

Remarkably, we found that the entire Gamilaraay nation would need to reduce to just 24 reproductive individuals for couples to be as closely related on average as full first cousins. This result assumes, however, that inbreeding coefficients at the great grandparent stage are negligible. While reasonable, this will of course not always be true. For this reason, we then dropped this assumption and asked instead how might kinship coefficients change over time. We found that kinship coefficients may both increase and decrease \emph{in reaction} to historical values, and changes in $n$. In particular, future values will decrease if in the past they were larger than a critical value which we label $f_{critical}$. Conversely, future values will increase if instead in the the past they were smaller than this $f_{critical}$. In this way, the Gamilaraay kinship trades-off the costs of being too closely related (such as incidence of recessive diseases \cite{wright1922coefficients, crow1986basic}) against the benefits of cooperative behaviours, accrued through increasing relatedness, as understood formally through Hamilton's Rule \cite{frank2012natural,gardner2011genetical}.

K Langloh Parker, in an early colonial account, remarked on this very system `the Blacks were early scientists in some of their ideas, being before Darwin with the evolution theory ... I rather think the Central Australians have the key to it. One old man here was quite an Ibsen with his ghastly version of heredity.' \cite{parker1905euahlayi}. In this way, there was some early recognition (missed almost entirely by early anthropologists) of the scientific underpinnings of these systems. For future work, it will prove fruitful to reanalyse other kinship systems whose utility has not yet been fully understood. In particular, taking a more probabilistic approach to gauge how relatedness is distributed in societies operating under a given system, instead of searching for broad kin restriction rules, will be insightful.   

\section*{Competing interests}
We have no competing interests.
\section*{Authors' contributions}
JMF carried out the research and wrote the manuscript. 

\section*{Acknowledgements} Gratitude is given to Garruu Winanga-li-Gii for his patience in learning about the Gamilaraay system.  The author is also grateful for  the hospitality and support  of the  Sydney Mathematical Research Institute (SMRI) during which many revisions were made.
\section*{Funding}
JMF is funded by a McKenzie Fellowship at the University of Melbourne.

\bibliographystyle{vancouver}
\bibliography{ref_2021}
\end{document}